\documentclass[journal]{IEEEtran}
\usepackage{cite}
\usepackage{amsmath,amssymb,amsfonts}
\usepackage{algorithmic}
\usepackage{graphicx}
\usepackage{textcomp}
\usepackage{hyperref}
\usepackage{balance}
\usepackage{soul}
\usepackage{adjustbox}

\begin{document}

\title{Phase Aberration Correction for \textit{in vivo} Ultrasound Localization Microscopy Using a Spatiotemporal Complex-Valued Neural Network}

\author{Paul Xing$^1$, Jonathan Porée$^1$, Brice Rauby$^1$, Antoine Malescot$^{2,3}$, Eric Martineau$^{2,3}$, Vincent Perrot, Ravi L. Rungta$^{3,4}$, and Jean Provost$^1$
\thanks{Corresponding author: jean.provost@polymtl.ca}\\
\and
$^1$Department of Engineering Physics, Polytechnique Montréal, Montréal, QC H3T 1J4, Canada
\\
$^2$Department of Physiology and Pharmacology Université de Montréal, QC H3T 1J4, Canada
\\
$^3$Department of Stomatology, Université de Montréal, QC H3T 1J4, Canada
\\
$^4$Centre Interdisiplinaire de Recherche sur le Cerveau et l’Apprentissage, Université de Montréal, QC H3T 1J4, Canada}

\maketitle

\begin{abstract}

Ultrasound Localization Microscopy (ULM) can map microvessels at a resolution of a few micrometers (µm). Transcranial ULM remains challenging in presence of aberrations caused by the skull, which lead to localization errors.  Herein, we propose a deep learning approach based on complex-valued convolutional neural networks (CV-CNNs) to retrieve the aberration function, which can then be used to form enhanced images using standard delay-and-sum beamforming. CV-CNNs were selected as they can apply time delays through multiplication with in-phase quadrature input data. Predicting the aberration function rather than corrected images also confers enhanced explainability to the network.  In addition, 3D spatiotemporal convolutions were used for the network to leverage entire microbubble tracks. For training and validation, we used an anatomically and hemodynamically realistic mouse brain microvascular network model to simulate the flow of microbubbles in presence of aberration. The proposed CV-CNN performance was compared to the coherence-based method by using microbubble tracks. We then confirmed the capability of the proposed network to generalize to transcranial  \textit{in vivo} data in the mouse brain (n=3). Vascular reconstructions using a locally predicted aberration function included additional and sharper vessels. The CV-CNN was more robust than the coherence-based method and could perform aberration correction in a 6-month-old mouse. After correction, we measured a resolution of 15.6 µm for younger mice, representing an improvement of 25.8 $\%$, while the resolution was improved by 13.9 $\%$ for the 6-month-old mouse. This work leads to different applications for complex-valued convolutions in biomedical imaging and strategies to perform transcranial ULM.

\end{abstract}


\section{Introduction}

\begin{figure*}[htp!]
\centering
\includegraphics[width=0.95\textwidth]{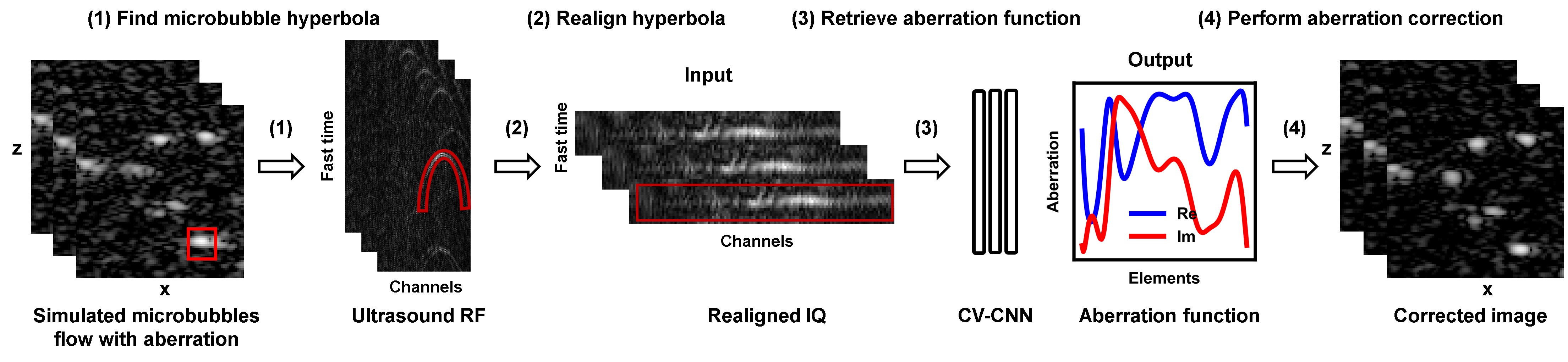}      
\caption{General framework for the phase aberration correction. Realistic microbubbles flow in presence of aberration and fully developed speckle are simulated. Step (1) The hyperbolas of the microbubble trajectory are retrieved from channel data. Step (2) Channel data are realigned around the microbubble hyperbola. Step (3) The realigned hyperbola is used to train the neural networks. The ground truth is the simulated aberration law which is retrieved by the neural network. Step (4) Correction is performed by reinjecting the aberration function into the beamformer.}
\label{fig:figure1}
\end{figure*}

Ultrasound localization microscopy (ULM) is a high-resolution imaging technique based on the localization of injected microbubbles over multiple frames\cite{viessmann2013acoustic, oreilly2013super, errico2015ultrafast}. ULM has been used for the imaging of the microvascular networks of various organs such as the brain, liver, and kidney at a resolution of a few micrometers (µm) \cite{couture2018ultrasound}. Such a high resolution is however sensitive to variations in transit time that can be caused by aberrations, which is particularly problematic for brain imaging where the aberrations caused by the skull can lead to inaccurate vascular maps that contain deformed, spurious, or missing vessels. This is even more problematic for dynamic ULM (dULM), which relies on higher microbubbles concentrations\cite{bourquin2021vivo}.

To limit phase aberration, ultrasound imaging of the brain is often performed through a thinned or removed-skull imaging window in animal models \cite{errico2015ultrafast, hingot2017subwavelength, deffieux2018functional}. However, this strategy is not compatible with clinical applications and limits pre-clinical studies in small animals. As vascular dysregulation is central to several neurodegenerative diseases and brain disorders, surpassing this limitation could lead to establishing new ULM-based biomarkers for brain diseases \cite{bourquin2021vivo,lowerison2022aging}. Multiple methods have been proposed in the literature for phase aberration correction \cite{ng1994comparative}. They include quality factor measurements \cite{nock1989phase}, correlation-based strategies \cite{o1988phase}, time reversal\cite{ montaldo2011time, osmanski2012aberration}, time delay applied to correct the transmit during data acquisition \cite{maasoy2005iteration}. However, most of these methods either assume that the aberration remains consistent in the whole field of view or are not adapted for plane wave imaging. Other recently proposed strategies for plane wave imaging require numerous transmits angles ($>$100) at the expense of frame rate \cite{lambert2020reflection, lambert2020distortion, lambert2022ultrasound, bendjador2020svd}. More importantly, phase aberration correction is not commonly performed in ULM since specific correction methods are lacking and needed to be performed over hundreds of thousands of images.

In recent years, deep learning has proved to be a powerful tool for analyzing medical images. However, most convolutional neural networks (CNNs) used in the context of deep learning are limited to real-valued data, which is an important limitation in ultrasound imaging where complex in-phase/quadrature (IQ) signals are widely used for demodulation and processing. To use complex-valued data, most CNNs split the real and imaginary parts into two real-valued independent channels, in separate inputs, or in separated CNNs \cite{hyun2020nondestructive, khan2020adaptive, milecki2021deep}, in which case most of the complex-valued algebra relations between real and imaginary parts are lost. Although complex-valued CNNs (CV-CNNs) \cite{trabelsi2018deep} are still relatively new in medical image processing, a few studies have showed their capability to outperform their real-valued counterparts  \cite{cole2021analysis, rawat2021novel,lu2022complex}.

Herein we propose a deep learning approach based on 3D spatiotemporal CV-CNNs to retrieve aberration functions by using both spatial and temporal information of the backscattering signals originating from microbubble flows. The advantage of learning aberration functions rather than correcting the image directly is that the improvements to the image are only associated to modifications to the delays used during reconstruction. We trained and validated our CV-CNN using an anatomically and hemodynamically realistic mouse brain microvascular network model \cite{damseh2018automatic, milecki2021deep, belgharbi2021anatomically} for simulating ultrasound data in presence of phase aberration both in transmit and receive. We compared the performance of our proposed CV-CNN with an adapted version of the coherence-based method by using the same channel signals from individual microbubble tracks. We then evaluated the capability of our network to generalize to transcranial \textit{in vivo} data of the mouse brain (n=3) by using a locally predicted aberration function.

\section{Methods}

\begin{figure*}[htp]
\centering
\includegraphics[width=0.95\textwidth]{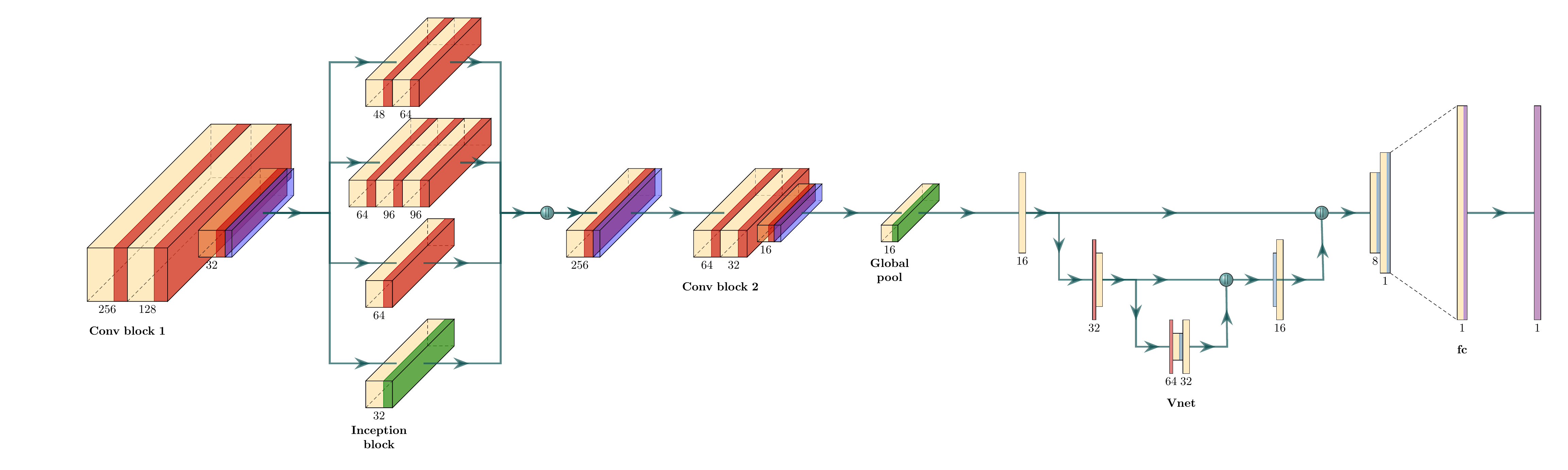}
\caption{Proposed CV-CNN architecture. The network consists of a first 3D convolution block followed by an inception block. The extracted features are then compressed into a 1D representation with a second convolution block combined with global pooling. Finally, a 1D V-net structure is used to compress the data into a lower representation, which is then upsampled to a 1D vector of the same dimension as the number of elements in the transducer probe. Numbers indicate quantity of channels in the CV-CNN. Fc: fully connected.}
\label{fig:figure2}
\end{figure*}

In this section, we first describe the simulation models we used to generate ultrasound IQ data in presence of phase aberration.  We then describe the general framework for our proposed CV-CNN (see Fig. \ref{fig:figure1}).

\subsection{Ultrasound simulation model}

A previously described \cite{damseh2018automatic, milecki2021deep, belgharbi2021anatomically}, graph-based models generated from the segmentation \cite{jegou2017one} of two-photon microscopy acquisitions of the vasculature of six mouse brains with Poiseuille flow conditions were used to simulate flow of microbubbles throughout the vascular network. Poiseuille flow is commonly assumed in ULM studies to model blood flow under normal physiological conditions either in simulation \cite{wiersma2022retrieving, heiles2022performance,lerendegui2022bubble} or for \textit{in vivo} quantification \cite{hingot2019microvascular, demeulenaere2022vivo,heiles2022volumetric}.

 Ultrasound radiofrequency (RF) data were simulated using a GPU implementation \cite{milecki2021deep,hardy2021sparse} of the SIMUS software \cite{garcia2022simus}. SIMUS was used to simulate the acoustic pressure coming from the simulated flowing microbubbles by using a 128-element, 15.625-MHz centered frequency ($f_c$) linear array transducer ultrasound probe with the properties of the L22-14v (Vermon, France) and 11 compounding angles (-5° to 5° with a 1° increment). The RF data were then subsampled into 100$\%$ bandwidth IQ channel data to emulate the \textit{Verasonics Vantage} system processing.

\subsection{Aberration simulation model}

We considered a near-field phase screen aberrator model that emulates a thin aberrator placed close to the probe. The aberration function in that case can be expressed as a complex-valued phasor
\begin{align}
y(n)=a(n)e^{i\omega \tau(n)},
\end{align}
where $\tau(n)$ stands for the time delays and $a(n)$ the amplitude attenuation for the $n^{th}$ element. A uniform distribution was used to generate random complex-valued aberrations (i.e., in amplitude and phase) for each element of the ultrasound probe. For purpose of this study, the phase aberration was limited to $\pm\lambda/2$ and the amplitude attenuation to 0-50$\%$. In mathematical form, this can be expressed as 

\begin{align}
a(n)& \sim U(0.5, 1)\\
\omega\tau(n)/2\pi& \sim U(-0.5, 0.5).  \nonumber  
\end{align}

 We then applied a cubic spline smoothing to have continuous aberration profile. The same aberrations were used for each compounding angle. The aberrations were then included in the simulated RF data both in transmit and receive.

\subsection{Complex-valued convolution}

CV-CNNs are still a new domain of study and are not yet fully supported with common deep learning libraries such as \textit{PyTorch}. We proceeded by using a customized implementation of the complex-valued representation of CNNs proposed by \cite{trabelsi2018deep}, which allows for the use of real-valued algebra to perform complex-valued computations. Other groups have made similar implementation of CV-CNNs to analyze ultrasound or magnetic resonance imaging (MRI) data \cite{lu2022complex, cole2021analysis, rawat2021novel}. In summary, the complex-valued convolutional kernel $W$ can be defined with a real ($W_r$) and imaginary ($W_i$) part, such as $W=W_r+iW_i$. By using complex algebra, the output $Y$ of such a convolution with a complex-valued input $X$ can be expressed as

\begin{align*}
Y&=(W_r+iW_i)*(X_r+iX_i)\\
&=(W_r*X_r-W_i*X_i)+i(W_r*X_i+W_i*X_r),
\end{align*}

which can be rewritten using real-valued matrix algebra as

\begin{align}
\begin{bmatrix}
Y_r\\
Y_i
\end{bmatrix}
=
\begin{bmatrix}
W_r & -W_i\\
W_i & W_r
\end{bmatrix}
*
\begin{bmatrix}
X_r\\
X_i
\end{bmatrix},
\end{align}

and preserves the relationship of the real and imaginary parts of a complex-valued convolution within the framework of real-valued convolution used by \textit{PyTorch}. One should note that the above expression is valid for any linear operator and can be applied for instance with complex-valued, fully connected layers. The use of complex-valued convolutions requires also to generalize batch normalization to complex numbers. Based on \cite{trabelsi2018deep}, the inputs normalization can be expressed as

\begin{align}
\tilde{X}=V ^{-1/2}(X-E[X])
\end{align}

where $E[X]$ is the expected value of $X$ and $V$ the covariance matrix of $X$

\begin{align*}
V
=\begin{bmatrix}
cov(X_r,X_r) & cov(X_r,X_i)\\ 
cov(X_i,X_r) & cov(X_i,X_i)
\end{bmatrix}.
\end{align*}

Finally, by using a polar expression, we can redefine weight initialization in the polar coordinates 
$W=\left|W\right|e^{i\phi}$, where $\left|W\right|$ follows a Rayleigh distribution and $\phi$ a uniform distribution between $\left[-\pi,\pi\right]$.

\subsection{Neural network model}

We used as input $\chi$ the aberrated realigned channel IQ data on a region of interest following a single microbubble used as target point
(representing a tensor of $N_\theta$ transmit angles, $N_t$ time samples and $N_e$ elements) over $N_f$ successive frames, i.e $\chi\in  \mathbb{C}^{N_f\times N_\theta\times N_t\times N_e}$. The time delays $t_n$ for rephasing the channel data for a microbubble at position $(x,z)$ was calculated by using the geometrical delays in the DAS beamformer \cite{montaldo2009coherent,perrot2021so}

\begin{align}
    t_n(x,z)= \frac{z\cos\theta+x\sin\theta+\sqrt{z^2+(x-x_n)^2}}{c},
\end{align}

where $x_n$ represents the position of each element of the transducer and $\theta$ the transmit angle. The advantage of this data representation is that the network has direct access to the temporal dimension of the channel data without having to learn the physics of delays computation associated with diffraction hyperbolas. Since \textit{PyTorch} can only compute convolutions up to three dimensions and our inputs are represented in four dimensions, we treated the compounding angles as the CV-CNN channels \cite{gasse2017high,lu2022complex}.

The network produces as output a prediction of the complex-valued aberration function, i.e., $h\left(\chi;\Theta\right)\in  \mathbb{C}^{N_e}$, were $\Theta$ are the CV-CNN parameters. We used a supervised approach to find the optimal parameters $\Theta$ mapping the inputs data with the true aberration functions $y\in  \mathbb{C}^{N_e}$ used as targets

\begin{align}
\hat{h}(\chi;\Theta) = \text{argmin }\sum_{(\chi,y\in D)} Loss(h(\chi;\Theta),y),
\end{align}

where $\hat{h}\left(\chi;\Theta\right)$ represents the optimal mapping. The rationale behind our CV-CNN is that rephasing channel data represents a mapping between a curved geometry (i.e., diffraction hyperbola) to a Euclidean geometry. Indeed, in absence of phase aberration, the rephasing return a straight line with respect to the channels. Although this problem seems simple when imaging a single scatterer, it becomes far more complicated when multiples scatterers and noise are present, such as with \textit{in vivo} data, where there will be crossings between the diffraction hyperbolas. It is then possible to extract features between the channel, time, and frame dimensions through complex-valued 3D convolutions. The first part of our CV-CNN consists of a succession of convolution blocks that compressed the 3D data into a lower dimensional representation (see Fig. \ref{fig:figure2}). We incorporated a 3D complex-valued inception-like block inspired by inception layers  \cite{szegedy2016rethinking, szegedy2017inception}. This strategy allows to deploy multiple convolution kernels and pooling sizes in parallel and then let the network extracts the appropriate features. The output of this inception-like block is then sent into another 3D complex convolution block with global pooling that compresses the data into a 1D representation. Finally, a simplified 1D V-Net-like block composed of a series of convolution and deconvolution blocks with skip connections to help regularizing the network is used to bring back the data dimension to the same size as the number of elements of the ultrasonic probe. The different kernel sizes and channel dimensions used are presented in Table \ref{tab:Kernel}. 

Complex batch-normalization was applied after each convolution. The complex ReLU (CReLU) \cite{trabelsi2018deep} was used as activation function, which is defined as

\begin{align}
CReLU(Y) = ReLU(Y_r)+iReLU(Y_i).
\end{align}

Although there are other implementations of complex activation functions, namely modReLU and zReLU, we chose to use CReLU because other studies on image classification and MRI reconstruction showed it outperformed all other complex activation functions  \cite{trabelsi2018deep, cole2021analysis}. The complex mean-squared loss (L2) function defined as

\begin{align}
L2= \frac{1}{N}\sum_{i=1}^{N} || \hat{h}(\chi_i;\Theta)-y_i||^2
\end{align}

was used for the learning along with the Adam optimizer  \cite{kingma2014adam} for the backpropagation. Dropout layers were added to regularize the learning  \cite{srivastava2014dropout}.

 \begin{table}[t]
    \centering
        \caption{Kernel sizes and channel dimensions}
    \begin{adjustbox}{width=1\linewidth}
    \tiny
\begin{tabular}{c|ccccc}
\hline
Block & Kernel sizes & Stride & Channels & Activation & Layer size\\
\hline
Input&&&3&& $N_\theta\times 16\times 17\times 128$\\
\hline
Conv block 1& (3,3,7)& &256& CReLU\\
 & (2,3,4)&2& 128& CReLU\\\
 & (1,1,1)&& 32& CReLU+Dropout &$32\times 4\times 4\times 64$\\
 \hline
 Inception & (1,1,1); (5,5,5) && 48; 64 & CReLU\\
 & (1,1,1); (3,3,3); (3,3,3) && 64; 96; 96 & CReLU\\ 
 & (1,1,1) && 64& CReLU\\
 & (3,3,3) && 32 & pool\\
 &&&&CReLU + Dropout & $256\times4\times4\times64$\\
 \hline
Conv block 2& (3,3,3) && 64 & CReLU\\ 
& (4,4,4) &2& 32 & CReLU\\ 
 & (1,1,1) && 32 & CReLU + Dropout\\ 
 &(3,3,3) && 16& Global pool& $16\times1\times1\times32$\\
\hline
V-net& Down: 3; 3 &2; 2& 32; 64 & CReLU\\
& Up: 3; 3 &2; 2& 16; 8 & CReLU & $8\times 32$\\
\hline
Deconv &3; 3&2; 2& 64; 128& CReLU \\
&&&&fully connected& $128\times 1$\\
\hline
\end{tabular}	
\end{adjustbox}
    \label{tab:Kernel}
\end{table}

\subsection{CV-CNN Training}

\begin{figure*}[htp!]
\centering
\includegraphics[width=0.95\linewidth]{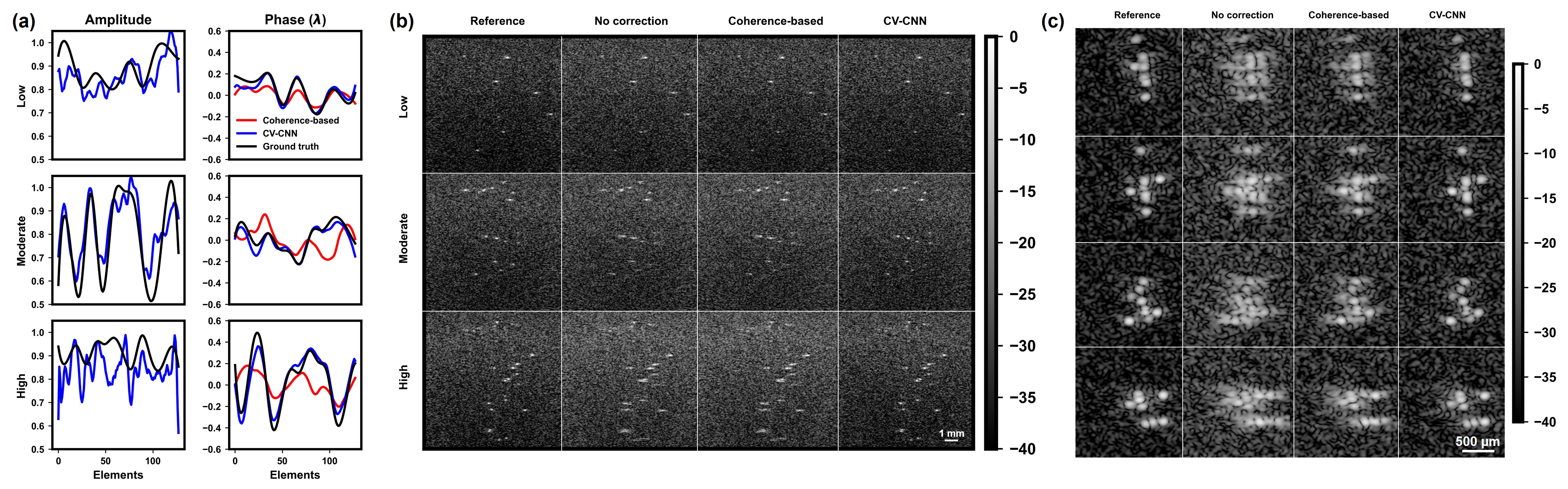}
\caption{Phase aberration correction performed on simulated microbubble flows. (a) The aberration functions retrieved by the CV-CNN are compared to the ground truth and the coherence-based method for different concentrations (low, moderate, and high). (b) Corrected images of microbubbles are compared to the coherence-based method and reference images for different concentrations. (c) Phase aberration performed on simulated overlapping microbubbles.}
\label{fig:figure3}
\end{figure*}

The general framework of our strategy is presented in Fig. \ref{fig:figure1}. \textbf{Step 1.}  Using our graph-model vascular networks, we simulated for each of the aberration functions the RF signals associated with microbubbles trajectories. We generated 20 000 different aberration functions that were then separated in a 4:1 ratio for the training and validation dataset, respectively. A separate test dataset consisting of 300 aberration functions was used neither during training nor validation. To emulate realistic RF data, we used a background of fully developed speckle (e.g., up to 50 point-sources contained in a $\lambda\times\lambda$ area) and electronic noise (5$\%$ Gaussian noise), which lead to a contrast ratio between the simulated point sources and the background of -25 dB after beamforming. We empirically found that a more robust model with increased capability to generalize to in vivo data could be achieved by using large fields of view (128$\lambda$ by 128$\lambda$) to ensure a variety of diffraction hyperbola eccentricities and varying concentrations (0.2-0.6 microbubbles/mm$^3$). The same aberration function was used for the whole FOV for each simulated RF data. We then proceeded by randomly selecting a single microbubble for each of these simulated RF to train the network. A different aberration function was used for each realigned hyperbola trajectory.

\textbf{Step 2.} We realigned the microbubble hyperbola using a 1/(4$f_c$) time sampling interpolation grid. The IQ input data size was set to $11\times16\times17\times128$, (corresponding to 11 compounding angles, 16 frames, 17 temporal samples and 128 elements). The time window used to realign the IQ dataset was chosen to contain the maximal possible phase aberration and the number of frames following the microbubble was set to 16.

\textbf{Step 3.} The realigned IQ were then used to train our neural network. The number of epochs during learning was set to 100 and the loss of the validation dataset was used to monitor the learning process. L2 regularization was used to avoid overfitting, with the regularization parameter $\alpha$ set to $10^{-4}$ and the batch size was set to 25. The initial learning rate was set to $10^{-4}$. An exponential scheduler was also used to adjust the learning rate during training with the decay rate $\gamma$ set to 0.99. After training, the independent test dataset was used to evaluate the performance of the CV-CNN. Both simulations and training were executed on GPU infrastructure (NVIDIA P100). Training required approximately 5 days to complete.

\textbf{Step 4.} Inference was performed on a computer with an Intel Core i5-9400 CPU (2.9GHz), 64 Go RAM, and a NVIDIA GeForce GTX 1080Ti GPU. Rephasing a microbubble track required 0.92s and inference 0.28s. Aberration correction was performed by including the time delays retrieved from the CV-CNN into the beamformer in receive and the average of the time delays in transmit.

\subsection{Coherence-based correction method}
We adapted the coherence-based correction method of  \cite{o1988phase} to evaluate the performance of the proposed CV-CNN. Briefly, hyperbolas of individual microbubble tracks were rephased as described previously. The relative phase delays were calculated by detecting the maximum of the cross-correlation between each element. A cubic interpolation was used to compute the delays associated to the maximum of each element. A low-pass filter composed of robust local regression smoothing was applied to the aberration profile to limit interference caused by noise and hyperbola overlaps. An average aberration function was then calculated for each track.

\subsection{Performance evaluations}

To evaluate the performance of the CV-CNN, we randomly generated an additional set of 100 individual microbubbles with phase aberration. Area under the curve (AUC) of the spatial coherence function between the elements after hyperbola realignment was used to quantify the performance of the phase aberration correction. For a $N$ element probe, the spatial coherence $R$ of a signal $s$ with respect to the lag $m$ is given by

\begin{align}
    R_m = \frac{N}{N-m}\frac{\sum_{n=0}^{N-m}\int_{-T/2}^{T/2}s_n(t)\cdot s_{n+m}^*(t) dt}{\sum_{n=1}^{N}\int_{-T/2}^{T/2}s_n^2(t)dt},
\end{align}

where $T$ represents the time window used for analyzing the signal and * the complex conjugate.

The shape of coherence function between element can be predicted by the van Cittert-Zernike (VCZ) theorem \cite{mallart1994adaptive}.  For an ideal point-like scatterer, the coherence is expected to be be constant since the echo should be in-phase for all elements after hyperbola realignment. For an ideal speckle, the coherence function is predicted as the autocorrelation of the aperture. For a linear probe, the spatial coherence of speckle is then given by a triangle function. The presence of phase aberration or partially-incoherent clutter noise will cause a deviation of the correlation function from its ideal curve \cite{long2020coherence}, leading to a reduced AUC. To evaluate the capability of the network to improve image quality, gain in contrast, lateral and axial width were used as metrics.

\subsection{Generalization to \textit{in vivo} Data}

\begin{table}[t]
\centering
\caption{Summary of phase aberration correction metrics}
    \begin{adjustbox}{width=1\linewidth}
\tiny
\label{tab:metrics}
\begin{tabular}{ccccc}
\hline
Metrics & Reference & No correction & Coherence-based & CV-CNN\\
\hline
Coherence (a.u.) & 0.73 $\pm$ 0.13 & 0.47 $\pm$ 0.17 & 0.54 $\pm$ 0.16 & 0.67$\pm$ 0.14\\
Contrast ratio (dB)& 24.7 $\pm$ 3.8 & 18.6 $\pm$ 5.5 & 20.6 $\pm$ 4.6 &22.2 $\pm$ 3.7 \\
Lateral width ($\lambda$) & 1.39 $\pm$ 0.16& 1.59 $\pm$ 0.45 & 1.41 $\pm$ 0.16 &1.38$\pm$ 0.16\\
Axial width ($\lambda$)& 0.85  $\pm$ 0.16 & 0.86 $\pm$ 0.16 &0.85 $\pm$ 0.15 &0.84 $\pm$ 0.14 \\
\hline
\end{tabular}
\end{adjustbox}
\end{table}

\begin{figure}[t]
    \centering
    \includegraphics[width=1\linewidth]{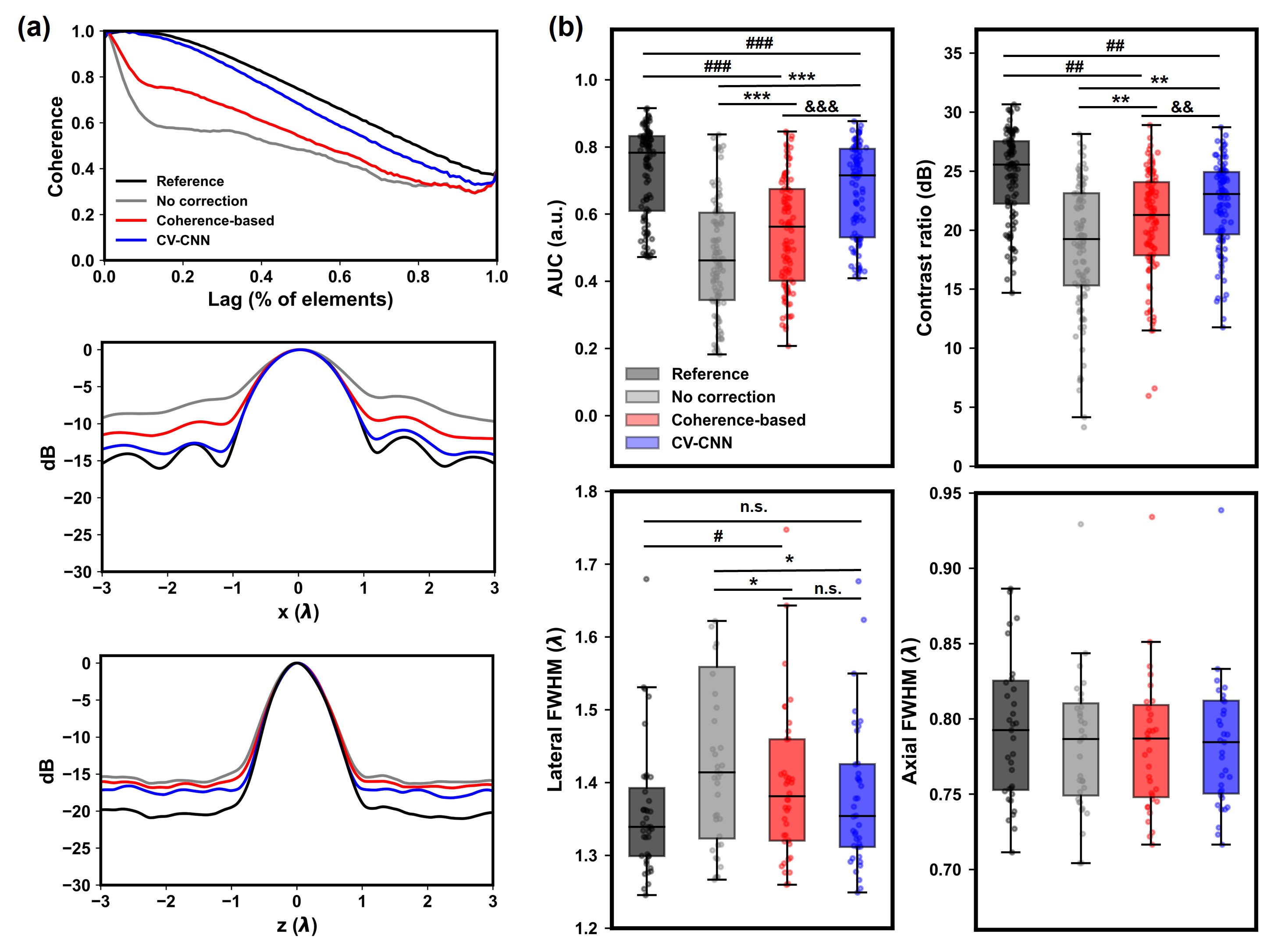}
    \caption{ Correction performance on simulated data. (a) Spatial coherence between elements along with the PSF of simulated microbubbles before and after phase aberration correction. In presence of phase aberration, correlation between elements is drastically reduced. The CV-CNN increased the autocorrelation to near reference value. The CV-CNN restores the PSF in both lateral ($x$) and axial ($z$) direction. (b) Comparison of AUC of the coherence function, contrast ratio, lateral and axial FWHM values for reference, aberrated, coherence-based method, and proposed CV-CNN with one-way ANOVA. \textit{Post hoc} one-tailed paired t-test statistically significance is marked with $\#$ for reference and correction comparison, with $*$ for aberrated and correction comparison, and with $\&$ for coherence-based method and CV-CNN comparison. AUC: area under the curve, FWHM: full width at half maximum.}
    \label{fig:figure4}
\end{figure}

\begin{figure}[t]
    \centering
    \includegraphics[width=1\linewidth]{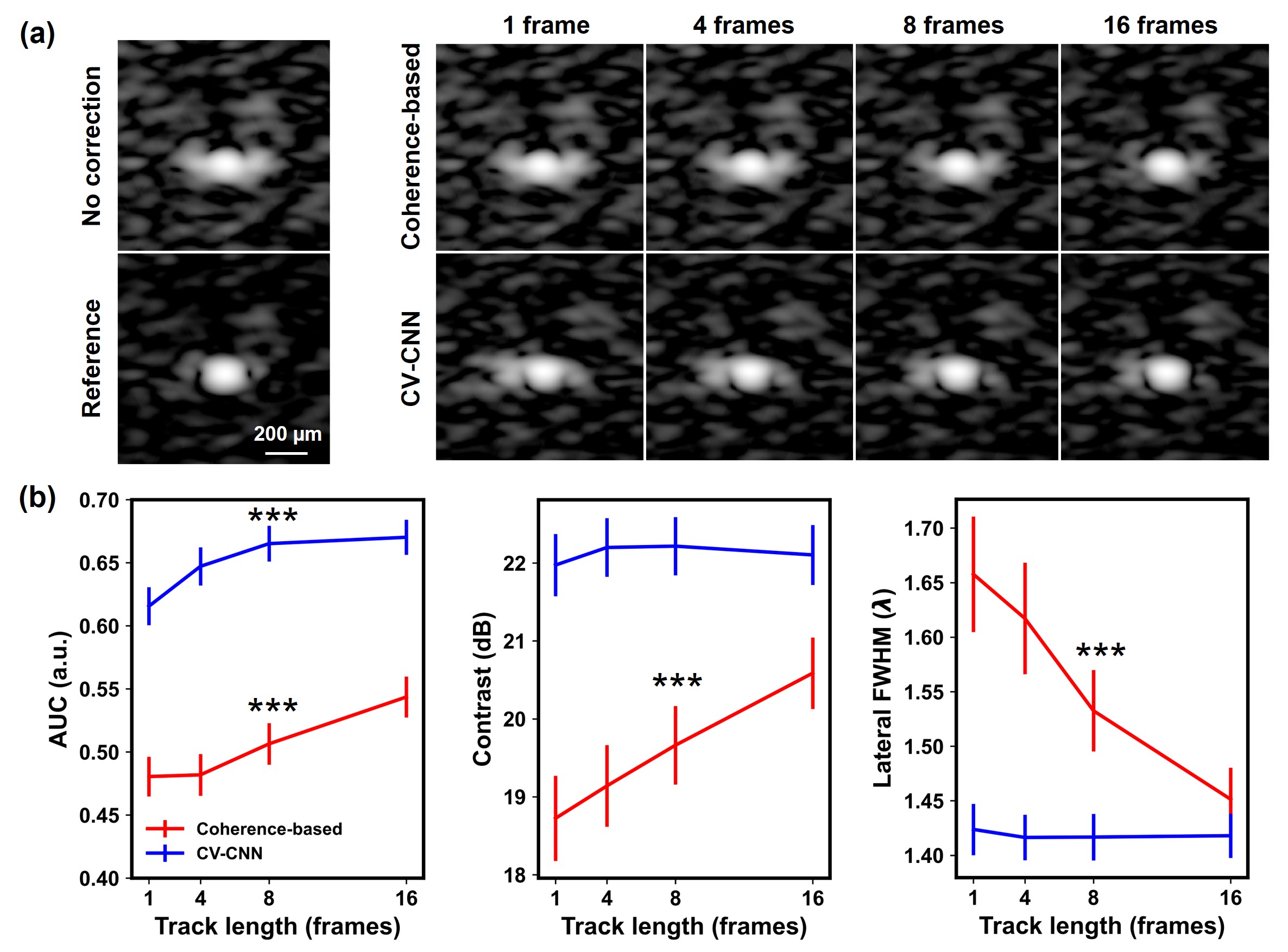}
    \caption{Effect of the number of frames on the performance of aberration correction on simulated data. (a) Aberration correction performed with aberration function retrieved for different track lengths. (b) Comparison of the AUC of the coherence function, contrast ratio and lateral FWHM for different track lengths. Statistical significance using a repeated measured ANOVA are marked with $*$. Increasing the number of frames improved the performance of both the CV-CNN and the coherence-based method.}
    \label{fig:figure5}
\end{figure}

Experimental procedures were approved by the Animal Care Ethics Committee of the University of Montreal (Permit Number: 21-017 and 22-013). Two 8-week-old male mice (mouse $\#$1 and mouse $\#$2) and one 6-month-old female mouse (mouse $\#$3) were used for \textit{in vivo} experiments. For each mouse, excess hair was removed from the head. The skin and skull were kept intact. A 3D-printed head mount was attached to the skull using cyanoacrylate glue, caudal to the imaging site, to limit movements from the head. For the vascular reconstruction dataset, we injected a bolus of 10 µL microbubbles (Perflutren Lipid Microsphere, Definity, Lantheus Medical Imaging, Billerica, MA, USA) in the retro-orbital veinous sinus. Raw ultrasound data were acquired with the \textit{Vantage system} (Verasonics Inc., Redmond, WA) in standard-frequency configuration at 100$\%$ sampling and a 15.625-MHz probe (L22-14, Vermon, France), with properties matching the simulations previously described. A complete acquisition lasted 5 minutes and consisted of 600 buffers composed of 500 images acquired at 1000 frames per second with 11 compounding angles (-5° to 5° with a 1° increment). The pulse shape was composed of three cycles of a sinusoidal wave with equalization pulses added at the start and end of each burst. The transmit voltage was 25 volts and mechanical index (MI) at elevation focus was 0.20. During acquisition, the mouse was kept under general anesthesia with Ketamine (50 mg/kg) and Medetomidine (1 mg/kg). The body temperature was maintained between 36 and 37°C.

Positions of microbubbles were first retrieved by applying a ULM pipeline as described previously \cite{milecki2021deep, bourquin2021vivo}. Briefly, microbubble signals were isolated from the tissue by using a singular value decomposition (SVD) clutter filter after beamforming \cite{demene2015spatiotemporal}. The SVD threshold was set heuristically to maximize removal of the tissue signal. Individual microbubbles were then detected by using correlation of their signal with the point spread function (PSF) associated to a single microbubble. To limit false detections, only microbubbles with correlation higher than 0.6 were considered. Detected microbubbles were tracked using the Hungarian method\cite{tinevez2019simple} with no gap filling, maximal linking distance of 2 pixels, and track lengths of 16 frames or more. Track positions were then used to realign the IQ for each detected microbubble. The rephasing was performed frame by frame to account for the change in position through time. 

In biological tissues, the aberration function varies within the entire field of view. Therefore, correction methods must be applied within the isoplanatic approximation. An aberration function was computed for every detected track and the mean position of each track was used to interpolate a map of the aberration function for each pixel point. A regularized smoothing algorithm based on the discrete cosine transform was applied to account for possible missing values in the aberration map \cite{garcia2010robust}.
For computing time considerations, only the last hundred buffers were used to determine the aberration correction function, which was then applied to the entire dataset. After manually adjusting the reconstruction speed of sound by qualitatively minimizing the lateral size of the microbubbles, the coherence-based approach or the CVCNN was applied to retrieve the aberration function. The local phase aberration function was incorporated into the beamformer as time delays to perform phase aberration correction. The ULM pipeline was then applied again on the newly detected microbubbles. To reconstruct the microvessel map, tracks were projected \cite{heiles2022performance} on a factor 10 interpolation grid. Only tracks longer than 25 frames were used for the ULM reconstruction. The same SVD and correlation thresholds were used before and after correction. The same tracks were used for comparison of the coherence-based method and the CV-CNN.

To evaluate the resolution before and after correction, the Fourier ring correlation (FRC) \cite{hingot2021measuring} was used by adapting code from \cite{diederich2018cellstorm}.  Images with higher resolution have a higher spatial frequency content and the FRC provides a measurement of the degree of correlation of two images at different spatial frequencies. It can be used as an estimate of the image resolution by splitting the localization dataset into two sub-images and establishing a threshold for the consistency of the frequency information. To generate the two sub-images, tracks used for calculating the density maps were randomly separated into two separate datasets. The FRC was computed by using the spatial spectrum $F_1$ and $F_2$ of each sub-image, such as

\begin{align}
    FRC(r) = \frac{\sum_{r\in R}F_1(r)\cdot F_2(r)^*}{\sqrt{\sum_{r\in R}|F_1(r)|^2\cdot \sum_{r\in R}|F_2(r)|^2}}.
\end{align}

The 1/2-bit threshold was used to assert image resolution, which corresponds to the highest spatial frequency containing at least 1/2 bit of information \cite{van2005fourier}. The bit threshold curve is based on Shannon theory, which stipulates that the rate of information I that can be transmitted by a signal with a given signal-to-noise-ratio (SNR) is $I=\log_2\left(1+SNR\right)$. For 1/2 bit in each sub-image, or a total information of 1 bit, this is equivalent to a SNR of 0.4142.

\subsection{Statistical analysis}

 For simulation data, a one-way ANOVA was used to assess difference between multiple groups. A \textit{post hoc} pairwise comparison using a one-tailed paired t-test with Bonferroni correction for multiple comparisons (p $<$ 0.01) was performed to evaluate statistical significance between two groups when the ANOVA was considered significant (p $<$ 0.05). A repeated measures ANOVA was used to assess statistical significance of the increase of the number of frames over the same metrics. For \textit{in vivo} data, a one-tailed Kolmogorov-Smirnov test was used to assert if one distribution was statistically larger than another (p $<$ 0.05).

\begin{figure}[t]
\centering
\includegraphics[width=1\linewidth]{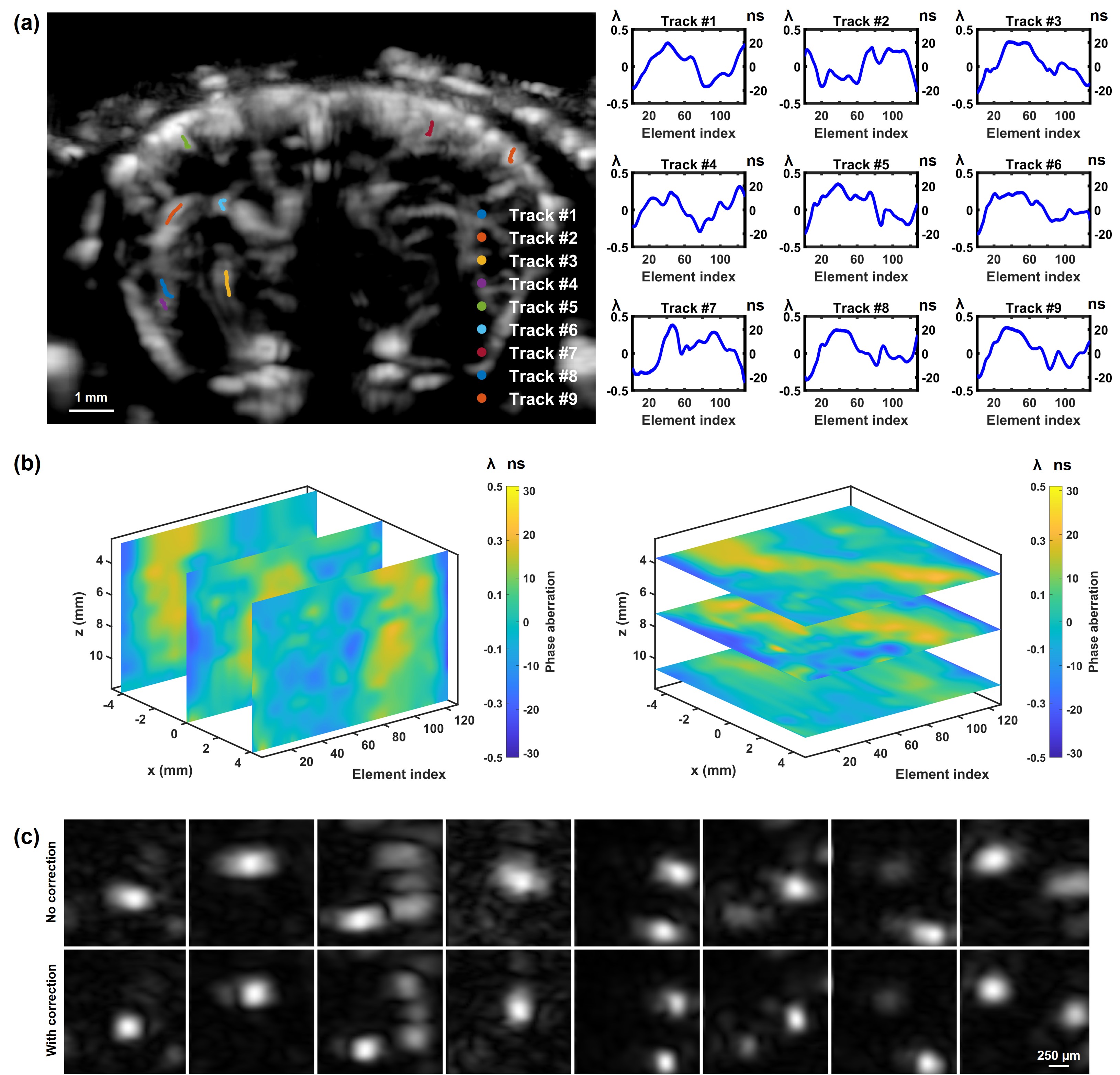}
\caption{\textit{in vivo} local aberration function in a mouse brain trough intact skull and skin. (a) Examples individual of aberration functions retrieved for different detected tracks. (b) Examples of the locally interpolated aberration function for different planes. (c) Examples of microbubble shapes without and with aberration correction. }
\label{fig:figure6}
\end{figure}

\section{Results}

We tested the capability of our network and of the coherence-based method to perform phase aberration correction for different microbubble concentrations on simulated data. While the coherence-based method could retrieve the aberration profile for low microbubble concentration, its performance degraded when the concentration was increased as shown in Fig. \ref{fig:figure3}a. The CV-CNN could retrieve the phase aberration for all microbubble concentrations used. However, a mismatch with the ground truth can be observed for the amplitude. Performing phase correction with the aberration function led to qualitatively enhanced images both in terms of resolution and contrast for all concentrations with the CV-CNN, while the coherence-based method only improved images with low microbubble concentrations as shown in Fig. \ref{fig:figure3}b. The presence of aberration distorted the shape of the simulated microbubbles, while corrected images showed microbubbles that resembled the reference images. Moreover, the CV-CNN could retrieve the aberration function when overlap between the microbubbles was present. Fig. \ref{fig:figure3}c shows that phase aberration correction could still be done in cases where the aberrated B-mode image included indistinguishable microbubbles, while the coherence-based method failed to do so.

\begin{figure*}[ht!]
\centering
\includegraphics[width=0.8\linewidth]{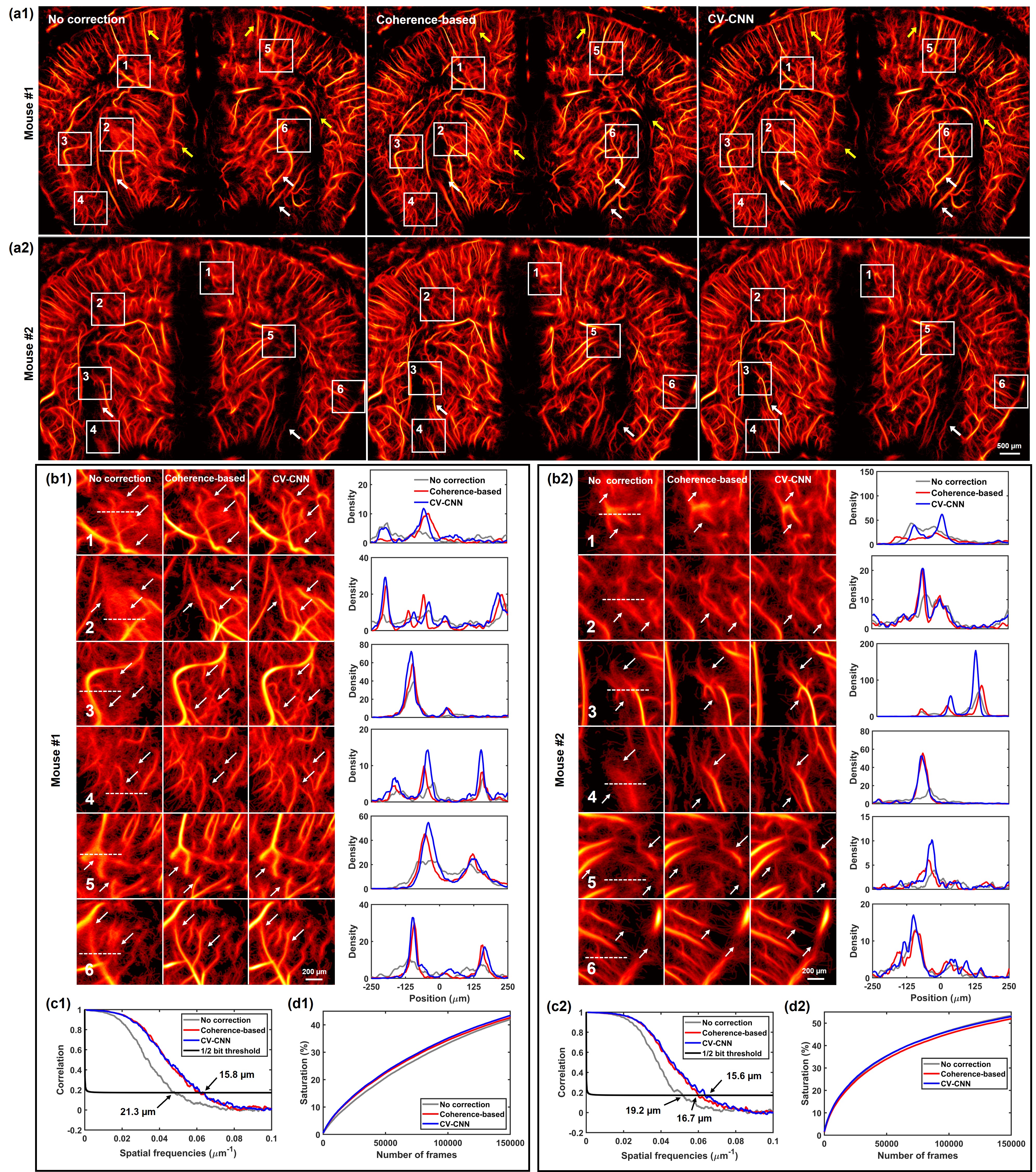}
\caption{\textit{In vivo} ULM reconstruction with aberration correction through intact skull and skin on two 8-week-old mice. (a1-a2) Vascular reconstruction without correction, with the coherence-based method and with the CV-CNN for mouse $\#$1 and $\#$2, respectively. (b1-b2) Microvessels extracted from regions indicated in a1 and a2. Profile comparisons show an increase in resolution after aberration correction. (c1-c2) Resolution measurement based on FRC for mouse $\#$1 before correction was 21.3 µm and improved to 15.8 µm after correction. Resolution for mouse $\#$2 before correction was 19.2 µm and improved to 15.6 µm after correction. (d1-d2). Saturation curve through ULM reconstruction was higher for the CV-CNN.}
\label{fig:figure7}
\end{figure*}

Fig. \ref{fig:figure4}a shows that in presence of the simulated clutter, the coherence function decreased with respect to the element lag. The presence of aberration led to a sharp deviation of the spatial coherence function, with coherence reaching below 0.6 after 10$\%$ spatial lag between elements. The CV-CNN correction led to an increase in the spatial coherence between elements that outperformed the coherence-based method, with a corrected curve similar to the reference one. Phase aberration is also associated with a degradation of both the lateral and the axial resolution, and with a decrease in contrast. Correcting the images with the CV-CNN or the coherence-based method both restored almost completely the PSF lateral width. Fig. \ref{fig:figure4}b shows that the area under the curve (AUC) of the spatial coherence function increased from 0.47 $\pm$ 0.18 to 0.67 $\pm$ 0.14 when using the CV-CNN, representing an improvement of 41.5$\%$ (p $<$ 0.0001). The correction was close to its reference value (0.73 $\pm$ 0.13), even if the difference remained statistically significant (p $<$ 0.0001). The CV-CNN performed significantly better than the coherence-based method (p $<$ 0.0001), which improved the AUC to 0.54 $\pm$ 0.16. The CV-CNN increased the contrast ratio by 3.6 dB (p $<$ 0.0001) from 18.6 $\pm$ 5.5 dB to 22.2 $\pm$ 3.7 dB and outperformed the coherence-based method (20.6 $\pm$ 4.6, p $<$ 0.0001). The CV-CNN significantly improved the lateral full width at half maximum (FWHM) from to 1.59 $\pm$ 0.45 $\lambda$ to 1.38$\pm$ 0.16 $\lambda$ (p $<$ 0.005) which was comparable to the reference value (p = 0.05). The coherence-based method also improved the lateral FWHM (p $<$ 0.005) and performed similarly to the CV-CNN (p = 0.03). The presence of aberration did not significantly impact the axial resolution (one-way ANOVA, p = 0.97). Those results are summarized in Table \ref{tab:metrics}. We then evaluated the effect of the number of frames on the performance of aberration correction. Fig. \ref{fig:figure5} shows that using a higher number of frames improved aberration correction for both the CV-CNN and the coherence-based method. Increasing the number of frames also significantly improved the AUC of the coherence function, contrast ratio and lateral FWHM for the coherence-based method (P $<$ 0.001), while only the AUC of the coherence function improved for the CV-CNN (p $<$ 0.0001).

The CV-CNN could also improve ULM performance on two 8-week-old adult mice. Fig. \ref{fig:figure6}a shows examples of different aberration functions for individual tracks for mouse $\#$1 and Fig.  \ref{fig:figure6}b shows the local aberration law after interpolation. Results show limited variation of the aberration function with respect the depth and more significant lateral variation. Notably, the phase aberration functions from the left side tend to be mirror images of the phase aberration functions retrieved on the right side of the brain, and aberration functions from the center regions of the brain showed a symmetric profile. After correction, microbubble shapes were more symmetrical as shown in Fig. \ref{fig:figure6}c.

\begin{figure}[htp!]
    \centering
    \includegraphics[width=1\linewidth]{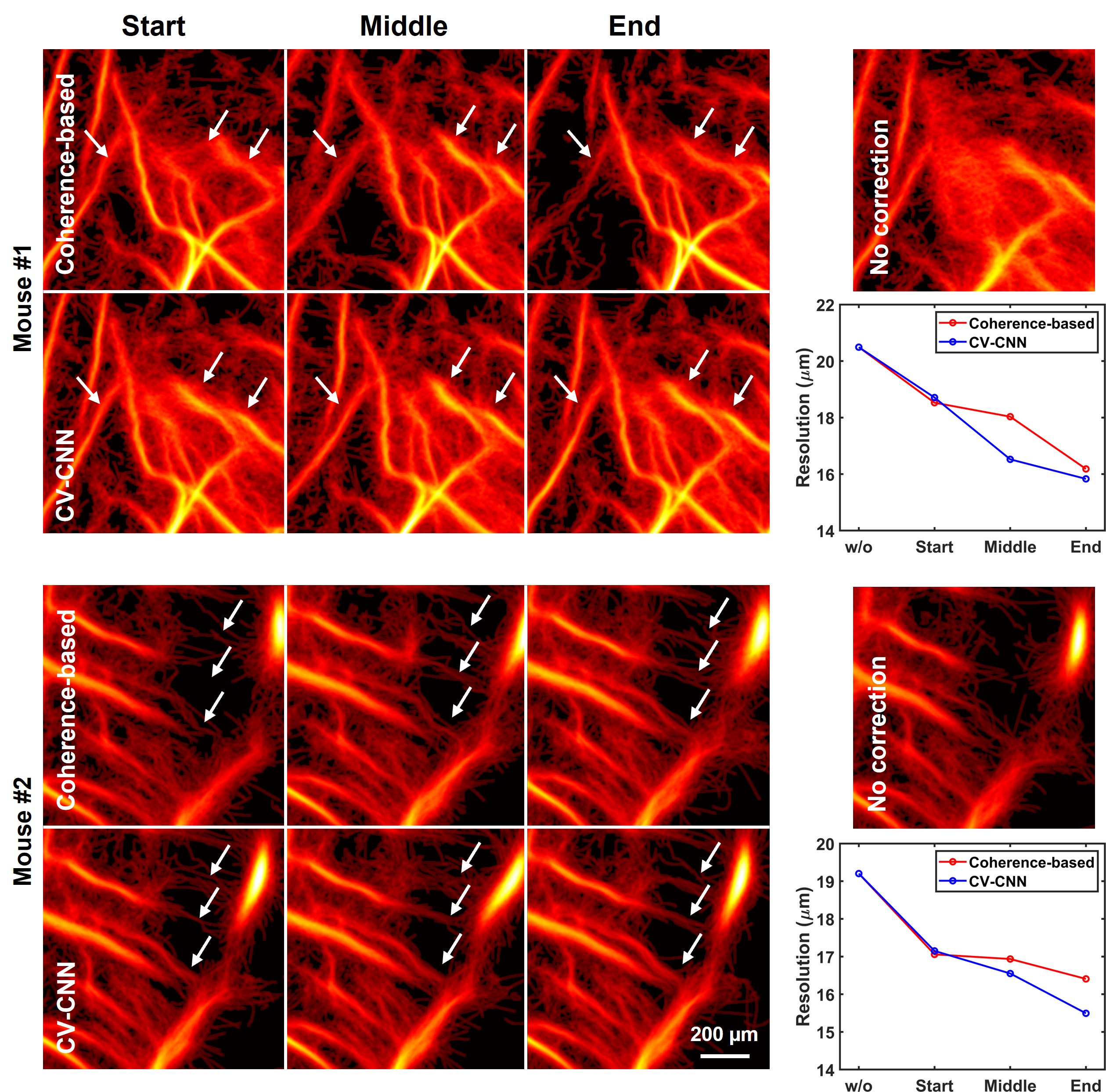}
    \caption{Effect of the microbubble concentration on the performance of aberration correction on \textit{in vivo} data. Aberration correction was performed by using buffers at the start (high concentration), middle, and end (low concentration) of the bolus injection. Resolution was measured with the FRC for the different concentrations. w/o: without correction.}
    \label{fig:figure8}
\end{figure}

ULM vascular reconstructions using the predicted aberration map improved resolution and contrast of microvessels Fig. \ref{fig:figure7}a shows that vessels appear sharper and more intense, and missing smaller vessels are now visible after correction. Overall, correction using the CV-CNN was more robust than the coherence-based method. White arrows indicate vessels that were separated in presence of aberration that were joined after correction by both methods, while yellow arrows indicate vessels that were deformed by the coherence-based method. We observed an increase of 24.8$\%$ of microbubbles density after correction with the CV-CNN for mouse $\#$1 and an increase of 11.8$\%$ for mouse $\#$2. To analyze the performance of the CV-CNN, we compared the profile intensities of selected microvessels in different regions of the mouse brain. Fig. \ref{fig:figure7}b shows an increase of microbubble density and reduced vessel width after correction for selected microvessels. Arrows indicate region where microvessels are sharper and better connected to adjacent vessels. Extracted vessel profiles also show that previously indistinguishable microvessels are now clearly separated after correction. The CV-CNN also shows a higher number of microvessels than the coherence based-method. Fig. \ref{fig:figure7}c-d shows the correlation of the spatial frequencies and saturation for the ULM reconstructed image of mouse $\#$1 and $\#$2. For both ULM images, the FRC curve moved to the right (in the direction of higher correlated frequencies) after correction with either the coherence-based method or CV-CNN. The FRC intersection with the 1/2-bit threshold curve established an overall spatial resolution of 21.3 µm and 19.2 µm before correction for mouse $\#$1 and mouse $\#$2, respectively. Resolution was down to 15.6 µm after correction for both the coherence-based method and CV-CNN, representing a gain of up to 25.8 $\%$. The saturation curve which counts the number of illuminated pixels during reconstruction\cite{heiles2022performance} showed that the CV-CNN performed better in detecting new structures. Fig. \ref{fig:figure8} shows that the CV-CNN was more robust than the coherence-based method at higher microbubble concentrations, with comparable or better resolution measured with FRC. 

\begin{figure*}[htp!]
\centering
\includegraphics[width=0.9\linewidth]{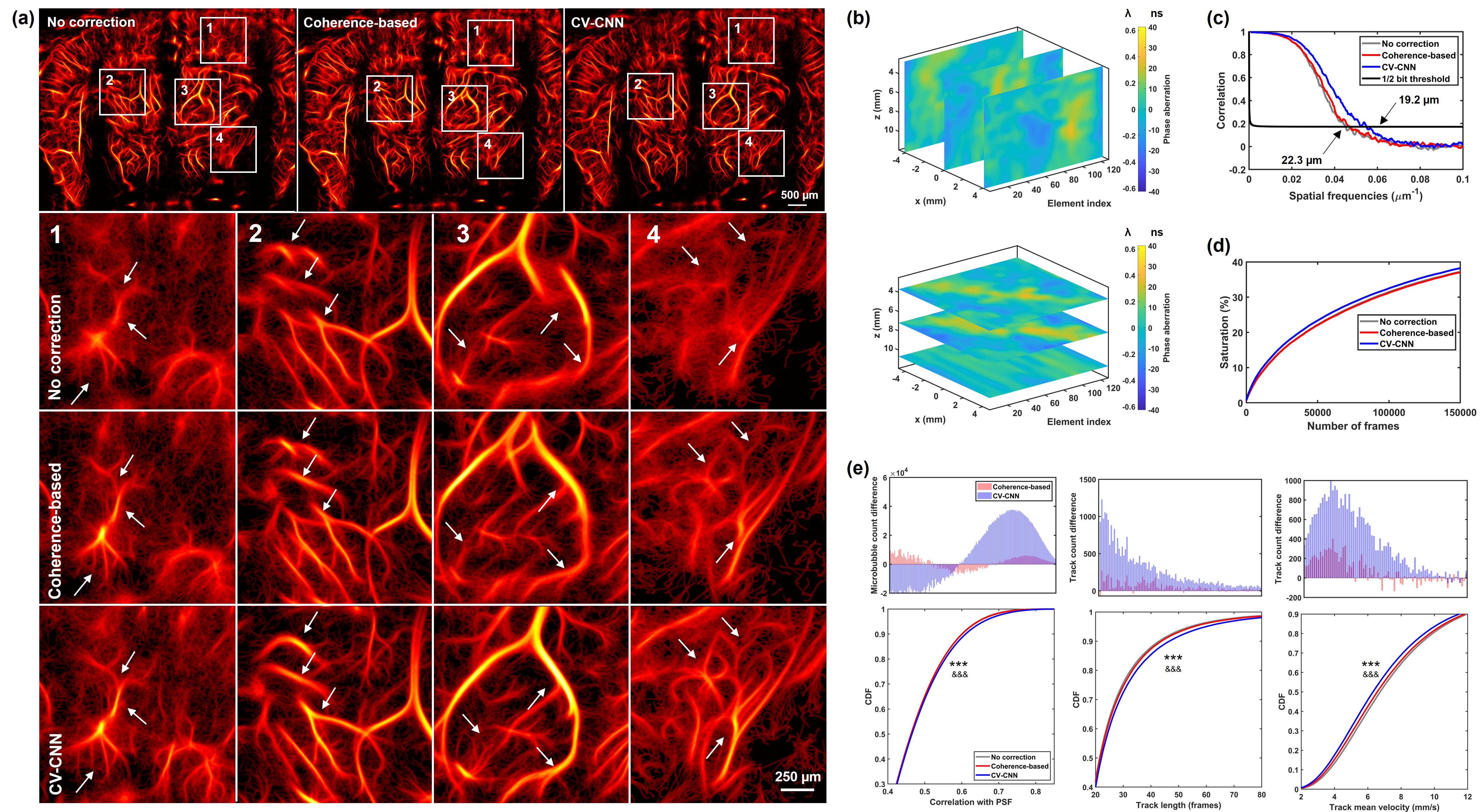}
\caption{\textit{In vivo} aberration correction through intact skull and skin in a 6-month-old mouse. (a) Vascular reconstruction before and after aberration correction using the coherence-based method and the proposed CV-CNN for mouse $\#$3. Correction using the CV-CNN showed a higher number of microvessels when compared to both the coherence-based method and in absence of correction. (b) Interpolated aberration function. (c)  Resolution measurement based on the FRC before correction was 22.3 µm and improved to 19.2 µm after correction with the CV-CNN. Correction with the coherence-based method did not improve the overall resolution. (d) Saturation curve was higher for the CV-CNN. (e) Comparison of microbubble correlation with PSF, track length and track mean velocity distributions for the coherence-based method and the CV-CNN. Microbubble and tracks counts are presented as difference from the distribution without correction. For CDF, statistically difference using the one-tailed Kolmogorov-Smirnov test is marked with $*$ for comparison between uncorrected and CV-CNN, and with $\&$ for comparison between coherence-based method and CV-CNN. CDF: cumulative distribution function, PSF: point spread function.}
\label{fig:figure9}
\end{figure*}

Fig. \ref{fig:figure9}a shows aberration correction in a 6-month-old mouse. Aberration correction using the CV-CNN reduced discontinuities and revealed additional vessels, while the coherence-based method had only limited impact on the overall image quality.  Shadowing effects are more prominent than in younger mice and remained after correction. The aberration map shows aberrations delays larger than the $\lambda$/2 used during training as shown in Fig.\ref{fig:figure8}b. The FRC curves shown in Fig.\ref{fig:figure8}c indicates an improvement in resolution after correction from 22.3 µm to 19.2 µm for the CV-CNN, while using the coherence-based method did not improve the resolution. The saturation curve shown in Fig.\ref{fig:figure8}d indicates that the CV-CNN performed better in reconstructing microvessels. The CV-CNN increased the number of microbubbles with high correlation, with a significant increase of the distribution when compared to both the coherence-based method and without correction (p $<$ 0.0001) as shown in Fig.\ref{fig:figure8}e. The CV-CNN also increased the number and length of tracks (p $<$ 0.0001). A larger number of tracks with lower mean velocities are detected after correction with the CV-CNN (p $<$ 0.0001).

\section{Discussion}

In this study, we proposed the use of a CV-CNN to directly retrieve the aberration function from realigned IQ data of microbubble flows to perform phase aberration correction in ULM. This strategy offers enhanced explainability in contrast with approaches where the output is the improved image, by ensuring image enhancements are associated only with aberration corrections. We first used simulated data of realigned microbubble hyperbolas to train our network. We then showed that it could correct for different levels of aberration and improved spatial coherence of simulated microbubbles by 41.5$\%$. Microbubble contrast, lateral and axial width were also restored to near reference values. We also showed that our proposed CV-CNN was more robust and outperformed the coherence-based method in every metric when stronger aberration was present. Including spatiotemporal information from microbubble tracks improved performance of both methods. Finally, we showed that our network could generalize to transcranial \textit{in vivo} data acquired on the adult mouse brain (n=3). FRC measurements on ULM reconstructed images after correction with a locally predicted aberration function  confirmed an improvement in microvessels resolution of 25.8 $\%$. Correction for stronger aberration was also demonstrated to be feasible for a 6-month-old mouse with thicker skull, while the coherence-based method failed to significantly improve image quality. We also showed that after correction with the CV-CNN, tracks with lower mean velocities are better detected, which is consistent with a better reconstruction of smaller vessels and increased resolution.

Developing correction methods are necessary for implementing clinical strategies for imaging the brain vascular network. Standard methods that exploit the presence of "guide stars" such as the coherence-based method required that the wavefront signals can be isolated. The lower performance of coherence-based method on \textit{in vivo} data could be explained by the fact that noise level and overlaps between hyperbolas are too important under normal experimental conditions for microbubbles to be used properly as guide stars. This is confirmed by our simulation results in which the coherence-based method performed poorly when the microbubble concentration was increased. To our knowledge, this is the first time a coherence-based aberration correction method is used in the context of ULM in the mouse brain. Hence, while we used this approach as a gold-standard and optimized it to the best of our ability, refinements could potentially further enhance its performance. The novelty of the CV-CNN approach resides in the combined use of spatiotemporal information of channel data and of a complex-valued network to encode the phase. The strength of CV-CNNs over conventional CNNs lays in the fact that inputs as well as outputs can be represented by complex numbers. This representation allows to preserve phase information and to consider amplitude as well as phase aberration within our model. Other studies using complex-valued CNN showed a better performance for retrieving phase information in biomedical images \cite{cole2021analysis}. 

The microbubble concentrations used during training could also reproduce situations when microbubbles can be used as target points and at the same time having various degree of diffraction hyperbolas overlaps. The presence of partially incoherent speckle noise also contributed to make the simulation more realistic. The CV-CNN had then to learn how to separate microbubble hyperbolas after IQ rephasing. This facilitated the capability of our network to generalize to \textit{in vivo} data, where crossing hyperbolas are inevitable with remaining clutter noise. The present framework also uses the positions of microbubbles detected by our ULM pipeline for the IQ rephasing. Our results showed the CV-CNN performed well even if the presence of phase aberration could have caused error on the initially detected positions. 

While \textit{in vivo} aberration profiles are not random and depend on physical properties between tissues, the CV-CNN performed well even when trained on randomly generated profiles. Using a substantial number of random aberration profiles allowed to cover a greater range of the aberration space and avoided possible bias coming from a less general model. The assumption was that this approach reduces the risk of overfitting since the network was not forced to learn a specific profile shape. By limiting the phase aberration function in our simulation model to $\lambda/2$ we were able to avoid phase unwrapping issues during the learning process, which is consistent with applications in mice at high frequency but also in humans at low frequency  \cite{demene2021transcranial}. Note, however, that even though the \textit{in vivo} data used in this study most likely included phase aberration beyond $\lambda/2$ as shown in Fig. \ref{fig:figure8}, the CV-CNN performed well. Those results suggest that even in presence of limited phase unwrapping, the relative change of the aberration phase between elements can still be retrieved with the proposed approach. Still, the CV-CNN approach would benefit from the combination of additional training with simulations that include larger phase aberration and robust phase unwrapping algorithms.

Since the aberration function is complex and the CV-CNN also returns a complex number, the method could potentially be used to correct both phase aberration and attenuation. Attenuation correction  could in principle, e.g., significantly improve the correction method for older mice, in which attenuation is more important because of the thicker skull. However, at this point our CV-CNN did not perform as well for retrieving the amplitude attenuation, even in simulations. We noticed that amplitude correction was more susceptible to added noise. On the other hand, the good performance of the phase correction highlights the ability of CV-CNNs to encode phase information from complex-valued data. Some complex activation functions like the CReLU are also better suited to discriminate the phase \cite{trabelsi2018deep}. Hence, development of novel activation functions that optimize both amplitude and phase would be highly beneficial for further applications of CV-CNNs. It is also known from previous work that increasing the number of transmit angles up to 100 improves the estimation of the aberration function \cite{lambert2022ultrasound,bendjador2020svd}. During validation of the CV-CNN, we indeed observed better convergence of the loss function when increasing the number of compounding angles, which lead us to use up 11 angles in our ULM pipeline, and it is possible that the CV-CNN would have performed even better if we had increased further the number of compounding angles in our \textit{in vivo} pipeline at the cost of reduced frame rate. The retrieved aberration functions had also important lateral variation. This could be explained by the fact that the skull cannot be adequately modeled by the phase screen aberrator because of the lateral variation in skull thickness. Those observations are consistent with aberration profiles observed in human patients \cite{demene2021transcranial,robin2022vivo} and emphasize the need for correction methods that can be performed locally or over multiple isoplanatic patches.

Limitations include aberration function and ultrasound simulation models that did not consider multiple reverberations, which can be present with \textit{in vivo} transcranial imaging. Considering multiple reverberations between the skull and brain would require simulating multilayered media, which could be done using, e.g., the K-wave toolbox  \cite{treeby2010k} but at the cost of longer computations. Moreover, the aberration function used remained the same for all compounding angles and the aberration correction was performed only in receive. The latter is not a limitation of the CV-CNN \textit{per se}, but of the DAS beamformer. Other image formation algorithms like the spatiotemporal matrix image formation (SMIF) \cite{berthon2018spatiotemporal} allows to perform correction both in transmit and receive and will be the object of future studies. Optimizing the local speed of sound\cite{robin2022vivo} or adjusting the start time of the acquisition by estimating the skull thickness could also have improved further the aberration correction performance. The current network architecture, i.e., the number of layers and kernel sizes, was based on using a predetermined and fixed input size. It is not clear how adapting the network for different input sizes, e.g., changing the number of temporal frames, would affect the performance on aberration correction. 

Current versions of common deep learning libraries such as \textit{PyTorch} and \textit{TensorFlow} do not allow to perform convolutions in more than three dimensions, which would become problematic when transposing the method to 2D matrix transducer array. Strategies like reshaping the aberration function into a 1D vector does not preserve properly local information and neighbor relationships. CV-CNNs are still in the developmental stage and not yet fully supported by common deep learning libraries and the customized implementation used herein could be suboptimal in terms of computational cost.

\section{Conclusion}
We demonstrated the capability of a CV-CNN to perform phase aberration correction by directly using IQ data. The CV-CNN could generalize to \textit{in vivo} transcranial ULM data and gives better resolved microvessels. Finally, the CV-CNN was also more robust than the standard coherence-based method. This work serves as a proof-of-concept that CV-CNN-based approaches are a viable solution for analyzing complex-valued representation of medical imaging data and represents a new step toward clinical application for transcranial ULM.

\section*{Acknowledgment}

This research was supported in part by the Institute for data valorization (IVADO), the Canada foundation for innovation, the Canada First Research Excellence Fund, the New Frontiers in Research Fund, the Canadian Institutes of Health Research (CIHR), Calcul Québec, and Compute Canada.

\bibliographystyle{ieeetran}
\balance
\bibliography{references}

\end{document}